\documentclass[]{raa}
\usepackage{deluxetable}
\usepackage{amssymb}            
\usepackage{graphicx,times}
\usepackage{natbib}
\usepackage{color}

\providecommand{\bm}[1]{\mbox{\bol
dmath$#1$\unboldmath}}

\begin{document}

\title{Photometric Calibration on Lunar-based Ultraviolet Telescope for Its First
Six Months of Operation on Lunar Surface}

\volnopage{}
	\setcounter{page}{1}

	\author{J. Wang	
	\inst{1}
	\and L. Cao
        \inst{1} 
        \and X. M. Meng
        \inst{1}       
        \and H. B. Cai 
        \inst{1}
        \and J. S. Deng
        \inst{1}
        \and X. H. Han
        \inst{1}
        \and Y. L. Qiu 
        \inst{1}
        \and F. Wang
        \inst{1}
        \and S. Wang
        \inst{1} 
        \and W. B. Wen
        \inst{1}
        \and C. Wu
        \inst{1}
	\and J. Y. Wei
	\inst{1}
        \and J. Y. Hu
        \inst{1}	
	}
 \institute{National Astronomical Observatories, Chinese Academy of Sciences,
             Beijing 100012, China; {wj@nao.cas.cn}
%
}

\abstract{
We reported the photometric calibration of Lunar-based Ultraviolet telescope (LUT), the first
robotic astronomical telescope working on the lunar surface, for its first 
six months of operation on the lunar surface. 
Two spectral datasets (set A and B) from near-ultraviolet (NUV) to optical band were constructed for 
44 International Ultraviolet Explorer (IUE) standards, because of the LUT's relatively wide wavelength coverage. 
Set A were obtained by extrapolating the IUE NUV spectra ($\lambda<3200\AA$) to optical band basing upon
the theoretical spectra of stellar atmosphere models.
Set B were exactly the theoretical spectra from 2000\AA\ to 8000\AA\ extracted from the same model grid.
In total, seven standards have been observed in 15 observational runs until May 2014. The calibration results 
show that the photometric performance of LUT is highly
stable in its first six months of operation. The  magnitude zero points obtained
from the two spectral datasets are also consistent with each other, i.e., 
$\mathrm{zp=17.54\pm0.09}$mag (set A) and $\mathrm{zp=17.52\pm0.07}$mag (set B).
\keywords{space vehicles: instruments --- telescopes --- techniques: photometric --- ultraviolet: general}}

 \authorrunning{Wang et al. }            
 \titlerunning{Photometry Calibrations on LUT}  
 \maketitle

\section{Introduction}

Lunar-based Ultraviolet Telescope (LUT) on board the Chinese first lunar lander 
(Chang'E-3) is the first robotic astronomical telescope
working on the lunar surface in the history of lunar exploration, and was
developed by National Astronomical Observatories of CAS (NAOC) and Xi'an
Institute of Optics and Precision Mechanics of CAS (XIOPM).  
Taking advantage of both very slow rotation and extremely tenuous atmosphere of the Moon, 
the main scientific goals of LUT are 1) to continuously monitor bright variable stars in the near-ultraviolet (NUV) band for as long as 
a dozen days; 2) to perform a dedicated sky survey at low Galactic latitude in the NUV band. The survey can 
cover the regions that have been avoided by the Galaxy Evolution Explorer (\it GALEX\rm) mission (Martin et al. 2005). We 
refer the readers to Cao et al. (2011) for a detailed description on the scientific goals and mission conception of LUT, 
and to Wang et al. (2011) for the discussion on the potential effect of the lunar exosphere on the 
NUV sky background emission detected by LUT. 

LUT was successfully launched on December 2nd, 2013 by a Long March-3B rocket, and 
had its first light on December 16th, 2013, two days after Chang'e-3 landed on the lunar surface. 
By May 2014, LUT has smoothly worked on the Moon for six months, and acquired a total of more than 50,000 images.
This paper describes the photometric calibration for its first 
\bf six \rm months of operation on the lunar surface.

\section{Instrument Overview}

A cross section of LUT is shown in Figure 1. The telescope is a F/3.75 Ritchey-Chretien system 
with an aperture of 150mm. A fixed flat mirror (the third mirror) is used to reflect the light to the Nasmyth focus to reduce the length of the 
whole instrument. A two-lens field corrector is located in front of the focal plane, and are used to correct the field curvature. 
A NUV coating is applied on the last lens. The transparency of the coating peaks at 
2500\AA, and has an effective width of 1080\AA. 
An UV-enhanced 1024$\times$1024 AIMO CCD E2V47-20 (manufactured by the e2v Company) operated in frame transfer mode
is chosen as the detector mounted at the Nasmyth focus. The CCD has a pixel size of 13$\mu$m,
and can be thermal-electrically cooled by as much as 40\symbol{23}C below its environment.
A rotatable flat mirror (the pointing flat mirror) with a size of 200mm$\times$166mm is mounted on a two-dimensional gimbal, which
is used to point and track 
a given celestial object. The gimbal has a designed pointing accuracy of 0.05\symbol{23}, which corresponds to about 38 pixels on
the focal plane.  
Two LED lamps with a center wavelength of 286nm are equipped to provide an internal flat field.
The filed-of-view of LUT is 1.36$\times$1.36$\mathrm{degrees^2}$, which corresponds to a pixel scale of 
4.76\symbol{125}$\mathrm{pixel^{-1}}$. The main characteristics of LUT are summarized in Table 1. 
A system flow chart of LUT was presented as the Figure 1 in Cao et al. (2011).

\begin{figure}[!hc]
\centering
\includegraphics[angle=0,scale=0.5]{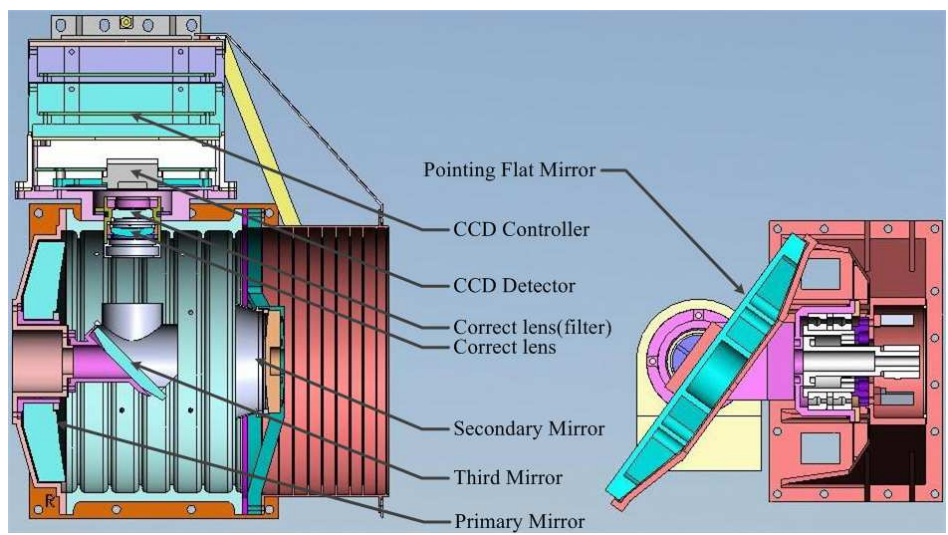}
\begin{flushleft}
\caption{A cross section view of LUT. The main components are labeled. }
\end{flushleft}
\label{Fig1}
\end{figure}

\begin{table}[!hc]
\caption{The LUT Prescription}
\centering
\begin{tabular}{ll}
\hline\hline
Item & Value \\
(1) & (2) \\   
\hline
Telescope:   &  \\ 
     Type             &  Ritchey-Chretien\\
     Primary diameter &  150mm\\
     Focal length     &  563mm\\
     Field of View    &  1.36$\times$1.36 degrees$^2$\\
     Peak efficiency  &  0.08 at 2500\AA\\ 
     Band width (FWHM)       &  1080\AA \\
  & \\
Pointing:    &  \\
     Pointing flat mirror  &  200mm$\times$166mm \\
     Gimbal           &  2-dimensions \\
     Gimbal accuracy  &  0.05\symbol{23} \\
  & \\
Detector: &    \\
     Type             & UV enhanced frame transfer AIMO CCD47-20\\
     Active pixels    & 1024$\times$1024\\
     Pixel size       & 13$\mu$m \\
     Pixel scale      & 4.76\symbol{125}pixel$^{-1}$ \\ 
\hline
\end{tabular}
\label{Tab3}
\end{table}

\section{Throughput Determination at Pre-launch}

The throughput of LUT at different wavelengths has been measured before launch in the laboratory by a 
dedicated calibration system. The calibration system is 
composed of a deuterium lamp 
(type L1314 by Hamamatsu, Inc.), 
a halogen-tungsten lamp (type L10296 by Hamamatsu, Inc.), a monochromator (type 207D by McPherson, Inc.), an adjustable diaphragm and
two pre-calibrated sensors. The telescope collects the collimated monochromatic light produced by the
monochromator without any intervening optics element on the light path.

Figure 2 presents the normalized throughput curve of LUT as a function of wavelength. The throughput peaks at around 2500\AA\ with 
a peak value of $\approx 8\%$.  The effective wavelength and width have been calculated under different definitions. 
The effective wavelength defined 
as the wavelength-weighted average is calculated to be 3046\AA. The effective wavelength defined by 
Schneider et al. (1983) 
\begin{equation}
 \ln\lambda_{\mathrm{eff}}=\frac{\int d(\ln\lambda)S_\lambda \ln\lambda}{Q}
\end{equation}
is calculated to be 2941\AA, where $Q=\int S_\lambda d\ln\lambda=0.027$ is the flux sensitive quantity and $S_\lambda$ the 
optical efficiency at wavelength $\lambda$. 
\rm The full width of half maximum (FWHM) of the curve is determined to be about 1080\AA. The rms fractional width $\sigma$ of the 
throughput, defined as 
\begin{equation}
 \sigma^2=\frac{\int S_\lambda\ln^2(\frac{\lambda}{\lambda_{\mathrm{eff}}})d\ln\lambda}{Q}
\end{equation}
is calculated \rm to be 0.146,

\begin{figure}[!hc]
\centering
\includegraphics[angle=0,scale=0.5]{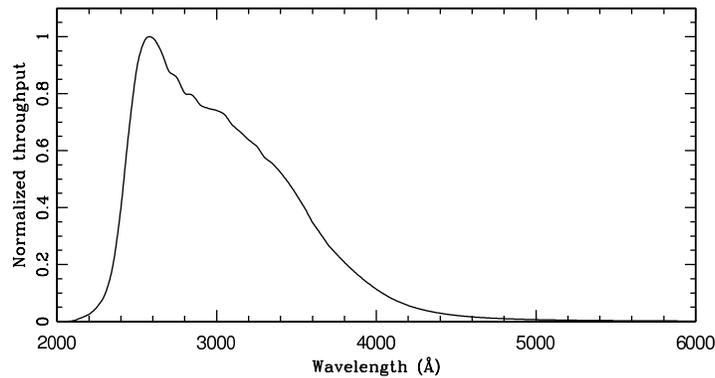}
\begin{flushleft}
\caption{The normalized throughput of LUT as a function of wavelength.}
\end{flushleft}
\label{Fig1}
\end{figure}

\section{Photometry Calibration}

\subsection{Magnitude System}

The AB magnitude system is adopted for LUT observations. 
This photometry system was first defined by Oke \& Gunn (1983), and has been widely used since its application in the Sloan Digital Sky Survey 
(Fukugita et al., 1996). It has also been used by the  \it GALEX \rm mission (Morrissey et al., 2005). 

The broadband AB magnitude of an object detected by LUT is defined as
\begin{equation}
  m_{\mathrm{AB}}=-2.5\log\frac{\int f_\nu S_\nu d\ln\nu}{\int S_\nu d\ln\nu}-48.6
\end{equation}
where $f_\nu$ is the specific flux density of the object in unit of 
$\mathrm{erg\ s^{-1}\ cm^{-2}\ Hz^{-1}}$ , and $S_\nu$ the LUT's total throughput at frequency $\nu$.
The throughput $S_\nu$ measured in the laboratory includes contributions from the optical efficiency of the telescope and 
pointing flat mirror, the transparency of the NUV coating, and the quantum efficiency of the detector.

The actual brightness $m_{\mathrm{LUT}}$ of an object observed by LUT is transformed from the observed instrumental magnitude $m_{\mathrm{inst}}$
through $m_{\mathrm{LUT}}=m_{\mathrm{inst}} + \mathrm{zp}$, where $\mathrm{zp}$ is the zero point of the LUT magnitude system  that 
should be determined by observing standard stars. 
Having the throughput $S_{\nu}$ measured, one can convert the observed magnitude to its physical flux
for a given object with known spectral shape.
In the conversion, there is no free parameter in the integrand of Eq.(3), except for a constant multiplier.

\subsection{Photometric Standard Stars}

The photometric calibration of LUT is based on the UV spectral atlas of the standard stars (Wu et al., 1998) provided by the 
International Ultraviolet Explorer (IUE) mission\footnote{
The IUE standard library can be found in the website http://www-int.stsci.edu/\~jinger/iue.html. There are in total
476 standard stars.}.  
This atlas was used because these stars 1) are bright enough for LUT to observe, 2) have reliable absolute specific flux in NUV, 
after the exclusion of the stars with known variability and peculiarity, 3) have known spectral types and luminosity classes (as well as  
the parameters of their stellar atmosphere), 4) have a comprehensive coverage on the H-R diagram, and 5) are nearly uniformly distributed on the sky.
The NUV spectra of all the IUE standard stars were taken in the low dispersion mode (LWP/R, 1850-3300\AA) with a spectral 
resolution of $\sim6\AA$. Their specific fluxes are uniformly obtained from the NEWSIPS pipelines, which involves the best 
treatment of noise and the last instrument parameters (Nichols \& Linsky 1996). 
The calibration of IUE satellite was based on the stellar atmosphere model of white dwarf G\,191B2B (Nichols \& Linsky 1996), 
which is the same as HST FOS. Although the uncertainty of the atmosphere model of G\,191B2B is only $\sim$2\%, 
the calibrated flux of IUE is systematically lower than that of HST by an amount of $\sim$6\%.  This is because
the model's effective temperature adopted by IUE is lower than that adopted by HST (58000 vs 63000 K).

There are in total 44 IUE standards located in the sky region available to LUT, as shown in Figure 3.
The total available sky area is about 3600degrees$^2$, which results in a surface density $\sim0.01\mathrm{degrees^{-2}}$ for the 44 standards.
Due to the scope of the gimbal rotation, LUT can only access a small sky area of $\sim400\mathrm{degrees^2}$ at a given time, which
means in average four standards are observable by LUT at a given time.
Such a surface density and the uniform distribution ensure that a photometric calibration can be done at any epoch and that the evolution of 
the LUT's efficiency can be continuously monitored. In addition, 
the model dependence of the final calibration results can be alleviated by statistics on multiple standards (see Section 4.5 below).


\begin{figure}[!hc]
\centering
\includegraphics[angle=0,scale=0.5]{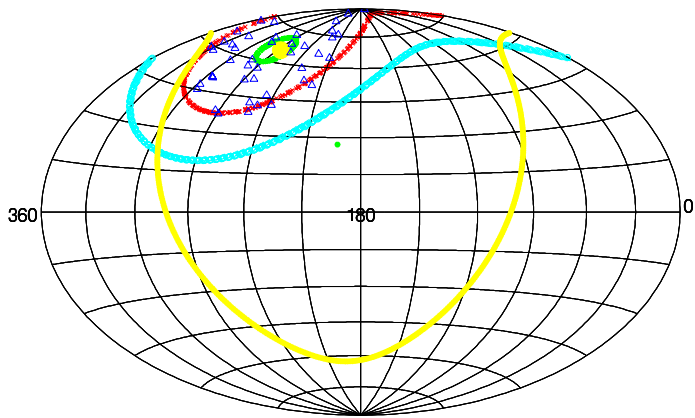}
\begin{flushleft}
\caption{The available sky of LUT and the distribution of the 44 IUE standards in the J2000 mean equator coordinate system. The 
LUT's available sky is located between the red and green circles. The locus of the zenith of LUT is shown by the 
cyan curve. The galactic plane and north pole of the Galaxy are shown by the yellow curve and small green dot, respectively,
The large yellow point marks the north pole of the Moon. The distribution of the standards is shown by the open blue triangles. }
\end{flushleft}
\label{Fig1}
\end{figure}

For the purpose of photometric calibration, two spectral datasets of the 44 IUE standards were constructed
to cover the wide wavelength range of the LUT throughput (Figure 2).
The spectral atlas provided by the IUE mission have a red cutoff at 3200\AA, which would result in an underestimate of the 
expected brightness if the original atlas were used in Eq(3). 
The level of the underestimation is expected to be more significant for red stars than for blue stars, 
simply because the peak of SED of a star shifts to long wavelength when the stellar surface temperature decreases. 
The ATLAS9 model atmospheres (Castelli \& Kurucz 2003) were adopted to solve this problem. With the theoretical spectra 
extracted from the model grid, we constructed two spectral datasets (set A and B) for the 44 IUE standards. 
Set A were obtained by combining the IUE observed NUV spectra ($\lambda<3200\AA$) with the theoretical spectra beyond 3200\AA. 
Set B were exactly the theoretical spectra from 2000\AA\ to 8000\AA.
We will show below that the two datasets return consistent calibration results within their uncertainties.



The theoretical spectra that we practically used were acquired as follows.
At first, a model spectrum was extracted from the original ATLAS9 mode grid by
a 3-dimensional linear interpolation for each of the 44 IUE standards. 
The three physical parameters used in the interpolation are the measured effective temperature,
surface gravity, and metal abundance that are collected from recent literature (see Table 2).
Secondly, each extracted model spectrum was reddened by its color excess $E(B-V)$ value 
through the IRAF.synphot.calcspec\footnote{IRAF is distributed by National Optical Astronomy Observatory, which is operated
by the Association of Universities for Research in Astronomy, Inc., under cooperative
agreement with the National Science Foundation.} task to account for the expected Galactic extinction.
The $E(B-V)$ value was taken from the IUE standard catalog, and the extinction curve from 
Calzetti et al. (1994). Finally, the 
absolute flux level of each reddened model spectrum is determined from the V-band magnitude reported in SIMBAD.
Figure 4 shows the so-constructed spectra (2000\AA\ to 8000\AA, both set A and B) of a standard star HD\,188665 (spectral type B5V) 
as an example. 

After the construction of the theoretical spectrum, two expected magnitudes (set A and B) in the LUT system were 
subsequently calculated through Eq. (3) for each of the standards.

\begin{figure}[!hc]
\centering
\includegraphics[angle=0,scale=0.5]{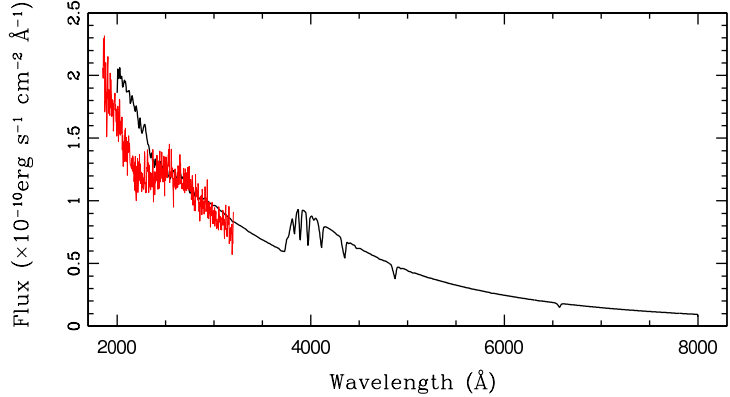}
\begin{flushleft}
\caption{The spectra of a standard star HD\,188665 in unit of FLAM. The black curve shows the spectrum extracted from 
the ATLAS9 stellar atmosphere model, and the red curve the NUV spectrum observed by IUE. The absolute flux level of the 
model spectrum is determined by its V-band brightness.}
\end{flushleft}
\label{Fig1}
\end{figure}

\subsection{Observations of the standards}

Seven standards have been observed until May 2014 in 15 observational runs. 
The details of the seven standards are listed in Table 2.
The effective temperature, surface gravity and abundance collected from the literature are tabulated 
in Columns (7), (8) and (9), respectively. Columns (10) and (11) list the
expected magnitudes for set B and A, respectively.
An appropriate observational strategy was designed to remove the strong stray light caused by the sunshine.
The telescope pointing was fixed with respect to the Moon in each observational run, which is required to be long enough, i.e., close to 30 minutes, to
allow the star to have a significant shift with respect to the fixed stray light pattern.
Each run is composed of a series of short exposures of 2 to 10 seconds depending on the brightness of the standards.
Thanks to the slow rotation of the Moon, the shift of a star on the focal plane within each exposure 
is negligible compared with the size of the point-spread-function (PSF).

\begin{table}[!hc]
\caption{IUE standards observed by LUT until May 2014.}
\centering
\tiny
\begin{tabular}{lccccccccccl}
\hline\hline
Star & s.p. type  & $\alpha$(J2000) & $\delta$(J2000) & $m_v$ & $E(B-V)$  & $\mathrm{T_{\mathrm{eff}}}$ & $\log g$ & Fe/H  & $m_{\mathrm{ck04}}$ &  $m_{\mathrm{iue}}$ & Reference \\
     &            &                 &                 &  mag  &   mag     &    K               &          &             &        mag           &   mag     &   \\
(1) & (2) & (3) &  (4) & (5) & (6) & (7) & (8) & (9) & (10) & (11) & (12)\\   
\hline
HD131873   &   K4III    & 14 50 42.3  &  +74 09 20 &  2.07 & 0.04 & 4077 & 1.7 & -0.10 &  6.52 & 6.52 & Koleva \& Vazdekis (2012)\\
HD153751   &   G5III    & 16 45 58.2  &  +82 02 14 &  4.23 & 0.00 & 5150 & 2.54 & 0.00 &  7.35 & 7.36 &  Soubiran et al. (2010) \\
HD159181   &   G2Ib-IIa & 17 30 26.0  &  +52 18 05 &  2.80 & 0.09 & 5325 &  1.51 & -0.02  & 6.09 &  6.08 & Koleva \& Vazdekis (2012)\\ 
HD164058   &   K5III    & 17 56 36.4  &  +51 29 20 &  2.23 & 0.01 & 3990 & 1.64 &  0.11  & 6.87 &  6.86  & Prugniel et al. (2011)\\ 
HD185395   &   F4V      & 19 36 26.5  &  +50 13 16 &  4.48 & 0.00 & 6700 & 4.30  &  0.01  & 6.34 &  6.41 & Cunha et al. (2000)\\
HD188665   &   B5V      & 19 53 17.4  &  +57 31 25 &  5.14 & 0.02 & 14893 & 3.86 & -0.17  & 5.26 &  5.33 & Fitzpatrick \& Massa (2005) \\
HD214470   &   F3III-IV & 22 35 46.1  &  +73 38 35 &  5.08 & 0.00 & 6637 & 3.59 & 0.09 & 7.18 & 7.22 &  Soubiran et al. (2010)\\
\hline 
\end{tabular}
\label{Tab3}
\end{table}

\subsection{Data reductions}

Reductions of the raw calibration data were performed with the general LUT reduction pipeline.
The pipeline is described in detail in a companion paper (Meng et al. in preparation). 
A brief summary of the pipeline is given below. 
At first, an overscan correction was applied on all the images taken by LUT. 
The overscan-corrected images in each observational run were then grouped according to the 
pointing of the gimbal with respect to the Moon. 
The total time elapsed in each group is required to be less than 30 minutes 
to avoid a significant evolution of both level and pattern of the underlying stray light.
In each group, the ingredient of stray light of a given image was acquired by combining all the images (except the given one) in the group without 
any image shift, in which a median value was extracted for each pixel.  
Before the combination, the images used to be combined were in advance 
shifted to have a common background level that was determined by an image statistics on the whole CCD of the given image. 
The underlying stray light was then removed from the given image by the constructed stray light image.
We emphasize that both underlying bias and dark current were simultaneously removed along with the 
stray light in this procedure.   
Each of the stray-light removed images was subsequently normalized by a composed flat filed.
In addition to the internal flat filed provided by the LED lamps that can only correct the nonuniform of CCD quantum efficiency in pixel scale,
a super flat filed was acquired by a drift observation of a normal star. 
As an example, Figures 5 and 6 show a finally used flat filed and a reduced image for HD\,188665, respectively.


The instrumental magnitude of each observed standard star was
obtained by an aperture photometry through the IRAF/apphot task. The aperture with a fixed radius of 7FWHM was used in
our photometry to ensure that almost all the signal from the star is enclosed in the aperture. 
The FWHM of each frame was determined by fitting the PSF by a moffat profile. 
The background level around the object was measured from an annulus with a width of 5 pixels out of the 
photometric aperture.

\begin{figure}[!hc]
\centering
\includegraphics[angle=0,scale=0.2]{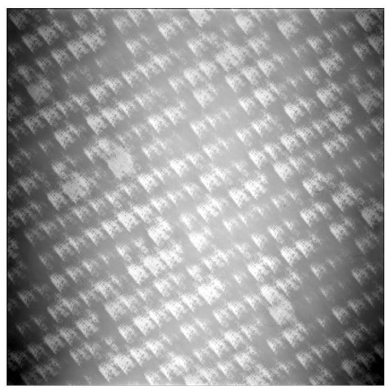}
\begin{flushleft}
\caption{A finally used flat filed derived from a combination of the internal flat field provided by the LED lamps and a super flat filed.}
\end{flushleft}
\label{Fig1}
\end{figure}

\begin{figure}[!hc]
\centering
\includegraphics[angle=0,scale=0.5]{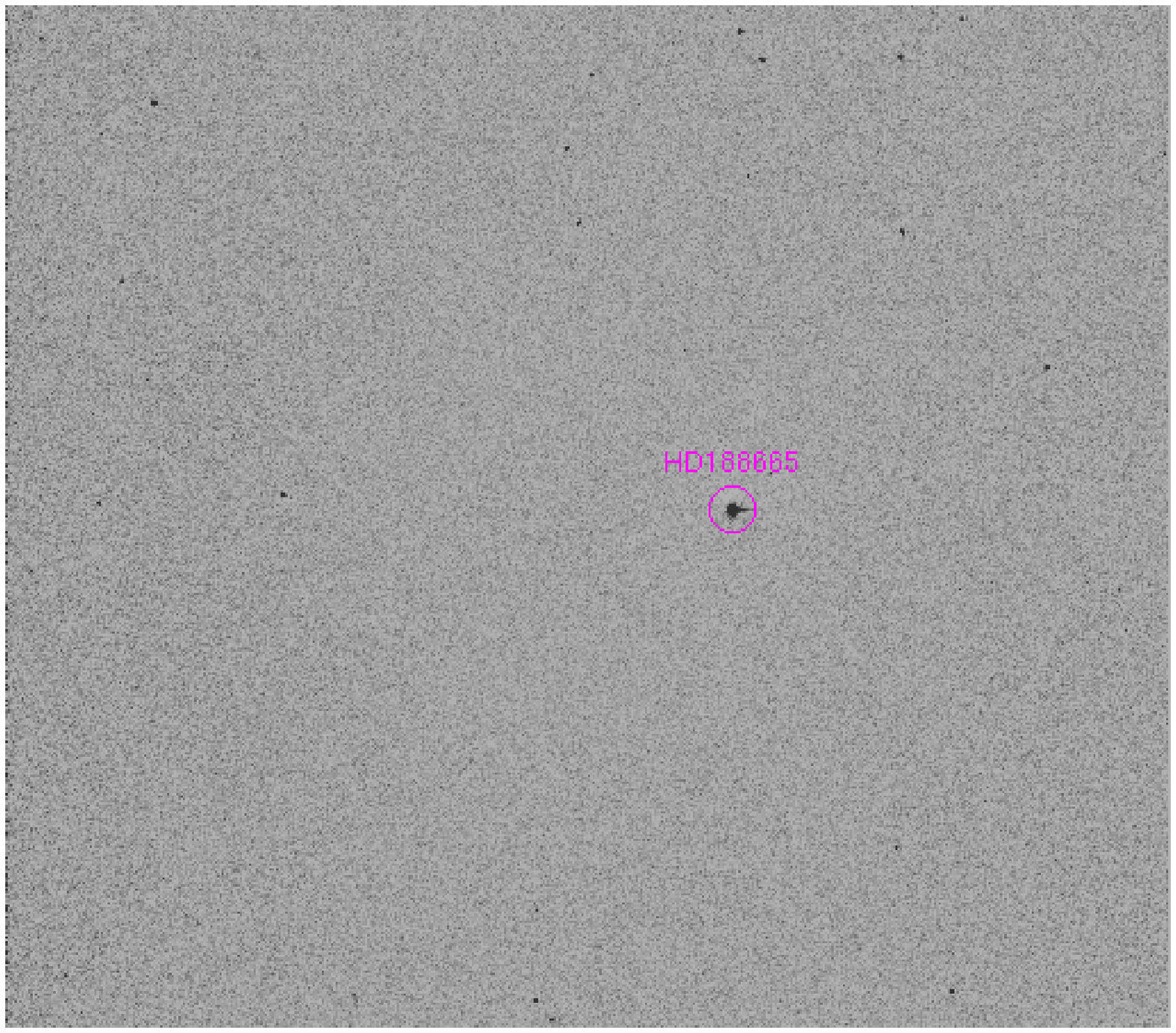}
\begin{flushleft}
\caption{A reduced image for standard star HD\,188665.}
\end{flushleft}
\label{Fig1}
\end{figure}

\subsection{Magnitude zero point}
 
The measured instrumental magnitudes with an arbitrary magnitude zero point $\mathrm{zp'} =25$mag are 
listed in Column (2) in Table 3. 
The $1\sigma$ errors reported in the column were derived from a statistics on the
multiple observations in each observational run. Each error contains the contributions mainly from the photon fluctuation, 
stray light removal and image normalization.
Because of their high signal-to-noise ratios, the uncertainty due to the photon fluctuation is generally as low as 
0.001-0.003 magnitudes for all the observed standards, which is much smaller than the reported errors. Our experiments show that 
a large fraction of the errors comes from the insufficient image normalization. As described in Section 4.4, the finally 
used flat filed is a combination of the intrinsic flat field and the super flat field. The 
intrinsic flat field can only normalize the nonuniform of CCD quantum efficiency in pixel scale, while the super flat field the 
nonuniform in a large scale of hundreds of pixels. 
The nonuniform in a middle scale of tens of pixels can therefore not be 
perfectly corrected by the combined flat field.

Column (3) in Table 3 lists the reduced zero points $\mathrm{zp'-zp}$ based on the expected magnitudes calculated from
the constructed spectral dataset of set B (i.e., Column (10) in Table 2). 
One can see from the table that three standards had been observed for 3-4 runs in LUT's 
first six months of operation. The resulted magnitude zero points of these three standards show non-detectable variation (within their 
uncertainties) with time, 
which indicates a highly stable performance of LUT in its first six months of operation.   

A statistic on all the calibration results from different standards at different time yields
two consistent average magnitude zero points obtained from the two constructed spectral datasets,
i.e., $\mathrm{zp=17.54\pm0.09}$mag (set A) and $\mathrm{zp=17.52\pm0.07}$mag (set B). 
Both reported errors correspond to a 1$\sigma$ significance level.
These reported errors are clearly larger than the errors inferred from 
a single standard. The increase of error can be easily understood because the expected magnitude of a standard depends on 
many factors, they are the IUE spectral/ATLAS9 model calibration (See Section 4.2 for the details), the V-band magnitude, the level of extinction and   
the parameters used to define the atmosphere model.

\begin{table}[!hc]
\caption{A comparison between the measured instrumental magnitudes and the expected ones.}
\centering
\begin{tabular}{lccc}
\hline\hline
Star &  $m_{\mathrm{inst}}$  &  $\mathrm{zp'-zp}$  & Epoch \\
               &       mag             &       mag           &       \\
(1) & (2) & (3) &  (4) \\   
\hline
HD131873    &  $14.16\pm0.03$   & 7.64 & 2014/05/19\\
HD153751    &  $14.75\pm0.02$   & 7.40 & 2014/05/19\\
HD159181    &  $13.51\pm0.01$   & 7.41 & 2014/01/12\\ 
            &  $13.52\pm0.02$   & 7.43 & 2014/01/16\\
            &  $13.55\pm0.01$   & 7.45 & 2014/02/10\\
HD164058    &  $14.40\pm0.02$   & 7.53 & 2014/01/12\\ 
            &  $14.44\pm0.30$   & 7.57 & 2014/02/10\\
            &  $14.37\pm0.01$   & 7.50 & 2014/02/15\\
            &  $14.42\pm0.03$   & 7.54 & 2014/03/16\\
HD185395    &  $13.79\pm0.01$   & 7.45 & 2013/12/23\\
            &  $13.79\pm0.01$   & 7.46 & 2014/02/15\\
            &  $13.82\pm0.01$   & 7.48 & 2014/02/18\\
            &  $13.78\pm0.02$   & 7.45 & 2014/03/16\\           
HD188665    &  $12.71\pm0.01$   & 7.45 & 2013/12/23\\
HD214470    &  $14.61\pm0.02$   & 7.44 & 2014/05/19\\

 \hline
\end{tabular}
\label{Tab3}
\end{table}

\section{Summary}

We have reported the photometric calibration of LUT for its first 
\bf six \rm months of operation on lunar surface. Two spectral datasets from 
NUV to optical band were constructed for photometric calibration 
for 44 IUE standards, because of the relatively wide wavelength coverage of LUT.
The calibration results indicate that the photometric performance of LUT is highly 
stable in its first six months of operation. The magnitude zero points inferred from the 
two spectral datasets are highly consistent with each other.




\begin{acknowledgements}

The authors thank the outstanding work of the LUT team and the support by the team of the ground system of CE-3 mission.
The study is supported by the Key Research Program of Chinese Academy of Sience (KGED-EW-603) and by the National Basic Research
Program of China (973-program, Grant No. 2014CB845800). JW is supported by
the National Science Foundation of China under Grant 11473036. MXM is supported by
the National Science Foundation of China under Grant 11203033.

\end{acknowledgements}

\label{lastpage}

\end{document}